\documentclass[sigconf,nonacm]{acmart}
\DeclareMathDelimiter{(}{\mathopen} {operators}{"28}{largesymbols}{"00}
\DeclareMathDelimiter{)}{\mathclose}{operators}{"29}{largesymbols}{"01}

\setcopyright{acmlicensed}
\copyrightyear{2025}
\acmYear{2025}
\acmDOI{10.1145/3744916.3787817}

\acmConference[ICSE '26]{The International Conference on Software Engineering}{April 2026}{Rio De Janeiro, Brazil}
\acmYear{2026}
\acmMonth{4}

\usepackage{enumitem}
\usepackage{amsmath}
\usepackage{graphicx}
\usepackage{microtype}
\usepackage{subcaption}
\usepackage{array}
\usepackage{tabularx}
\usepackage{multirow} 
\usepackage{algorithm}
\usepackage{algorithmic}
\usepackage[normalem]{ulem} 

\newcommand\blfootnote[1]{%
  \begingroup
  \renewcommand\thefootnote{}\footnote{#1}%
  \addtocounter{footnote}{-1}%
  \endgroup
}

\copyrightyear{2026}
\acmYear{2026}
\setcopyright{cc}
\setcctype{by}
\acmConference[ICSE '26]{2026 IEEE/ACM 48th International Conference on Software Engineering}{April 12--18, 2026}{Rio de Janeiro, Brazil}
\acmBooktitle{2026 IEEE/ACM 48th International Conference on Software Engineering (ICSE '26), April 12--18, 2026, Rio de Janeiro, Brazil}
\acmPrice{}
\acmDOI{10.1145/3744916.3787817}
\acmISBN{979-8-4007-2025-3/2026/04}

\renewcommand{\footnotetextcopyrightpermission}[1]{%
  \footnotetext[0]{\textit{Preprint. Accepted at ICSE 2026 -- 2026 IEEE/ACM 48th International Conference on Software Engineering, April 12--18, 2026, Rio de Janeiro, Brazil. DOI: \href{https://doi.org/10.1145/3744916.3787817}{10.1145/3744916.3787817}}}%
}
\begin{document}

\begin{sloppypar}

\title{TARIPlay: A Test Framework for AR Applications based on Interactive Area Tracking in Playback Videos}

\author{Sayed Amir Mousavi and Xiaoyin Wang}
\affiliation{%
  \institution{Department of Computer Science, University of Texas at San Antonio}
  \city{San Antonio}
  \state{Texas}
  \country{USA}
}
\email{{seyedamir.mousavi, xiaoyin.wang}@utsa.edu}

\renewcommand{\shortauthors}{Amir et al.}

\begin{abstract}



As Augmented Reality (AR) becomes more and more embedded in daily life, ensuring the quality, safety, and reliability of AR applications is increasingly important. However, AR apps present unique challenges for automated testing. Unlike static GUI layouts in traditional mobile apps, AR apps acquire their interaction interface from the surrounding environment, which is volatile and non-deterministic. Recent advancements like ARCore Playback and ARKit Replay allow developers to reuse real-world scenarios by recording and playing back enriched videos, enabling more feasible automated AR testing. However, using playback videos introduces two major challenges: test inputs must be timed precisely, and interactive areas in the video are dynamic, irregular, and difficult to identify. To address these challenges, we propose \textsc{TARIPlay}, a framework that analyzes playback videos to detect, track, and filter proper interactive areas over time for automated testing. In particular, \textsc{TARIPlay} identifies viable test opportunities based on criteria like stability and visibility, then feeds this information to an automated testing engine to simulate user interactions. We perform an experiment with four open-source AR apps and nine playback videos. Evaluation results show that TARIPlay significantly outperforms the existing tool Monkey in test coverage (55.8\% over 41.98\% on branch coverage) of AR-related code, and can also be used to assess the quality of playback videos for testing suitability.




\end{abstract}




\maketitle

\section{Introduction}

Augmented Reality (AR) overlays digital content onto real-world views in real-time for immersive user experiences. Applications such as \textit{Pokémon Go}~\cite{pokemon}, \textit{Amazon}~\cite{amazon}, \textit{IKEA Place}~\cite{ikea}, and \textit{Google Lens}~\cite{googlelens} demonstrate AR’s transformative potential across gaming, education, healthcare, and retail~\cite{Parekh2020}. As AR apps get more and more involved in daily activities, their quality, safety, and reliability attract more and more attention. Due to AR apps’ real-time interactions with complex environmental contexts in the real world, their bugs may cause more severe consequences, even immediate safety risks. For instance, in AR navigation applications, when imprecise content placement or improper occlusion handling are performed in the code, virtual elements may occlude critical aspects of the physical environment, leading to compromised user decisions and immediate risk for traffic accidents~\cite{10.1007/978-3-030-22666-4_12}. Similarly, a flaw in an AR-supported surgery system may lead to severe harm to patients if giving inaccurate guidance~\cite{kim2017virtual}. 

These real-world implications underscore the importance of robust testing and quality assurance practices in the development of AR apps. Despite the demand of high software quality, existing testing support for AR apps is still at its early stage. Traditional automated GUI testing frameworks fall short due to their inability to handle AR apps' interactions with real-world scenes~\cite{Minor2023}. As a result, AR developers often have to rely on manually constructed testing scenes and manual test execution~\cite{Minor2023}, which are labor-intensive and inefficient. Simulated environments such as Unity Mars~\cite{unitymars} have been recently developed, but it not yet clear how effective virtual reality scenes may simulate real-world scenes in AR testing, because they lack the noises and imperfections of the reality (e.g., coarse surfaces, diffuse reflections).



Recently, ARCore~\cite{arcore} and ARKit~\cite{apple2025arkit} both introduced recording and playback capabilities (i.e., ARCore Playback~\cite{playback} and ARKit Replay~\cite{replay}) to enable scenario reuse. These new features allow mobile phone users to record a video (enriched with sensor information) of the real world and run AR applications directly on the video. 
With these new features, the automatic execution of test cases for AR apps finally become feasible, and datasets of playback videos (e.g., ARTBank~\cite{artbank}) have been constructed to facilitate AR app testing. 

\begin{figure}
    \centering
    \includegraphics[width=0.9\linewidth]{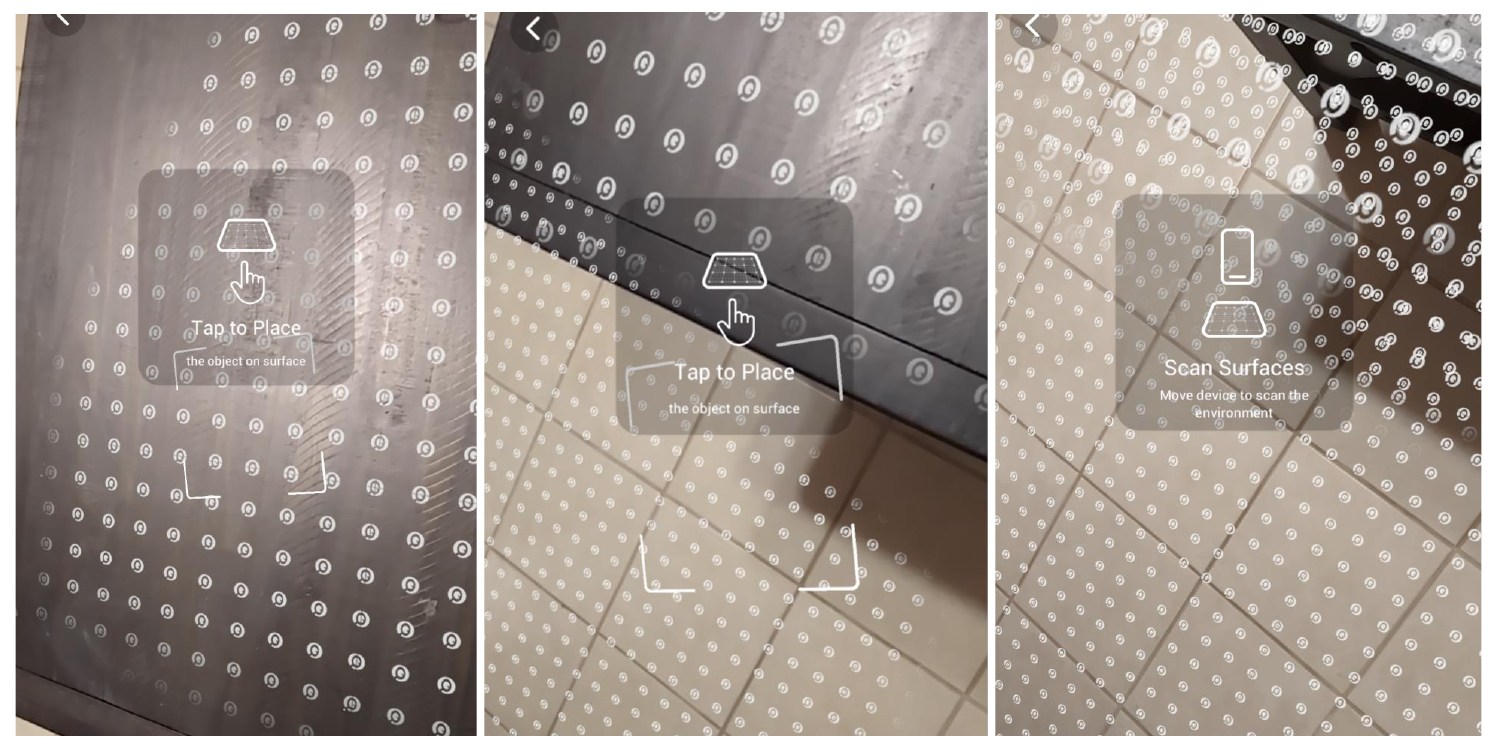}
    \vspace{-0.4cm}
    \caption{Dynamics of Planes in 2 seconds}
    \label{fig:playback}
    \vspace{-0.8cm}
\end{figure}

However, testing with playback videos is a setting much different from GUI testing~\cite{gu2019practical,qin2019testmig,liu2024make}, raising two novel challenges for automating the testing process. First, in automatic GUI testing, the testing process is fully controlled by the driver and we only need to feed a sequence of GUI events in order without worrying much about the time gap between them. But playback videos are strictly time-constraint, so test opportunities only occur at certain time period of a video. For example, a detected plane may show up at the 10$^{th}$ second of the video and disappear at the 21$^{st}$ second of the video. So an input event must be fed within the that time period to trigger a related app behavior. Second, unlike traditional GUI with predefined UI elements (e.g., buttons, checkboxes) and layouts, interactive areas (e.g., planes, human faces, collectively called \textit{trackables}) in a playback video are dynamically detected by the underlying AR framework so both their locations and bounding boxes are rapidly changing over time. Even worse, some interactive areas in the video may be of irregular shapes (e.g., a very narrow strip), moving fast on the screen, or disappearing quickly, so they may not be proper interactive areas for testing (and human users typically do not interact with them). To illustrate the two challenges, in Figure~\ref{fig:playback}, we show the dynamics of a detected plane over 2 seconds in a playback video. In the figure, the table top plane moves and disappears quickly and may not be a good interactive area to target in testing. Therefore, without knowing the time period, location, and size information about proper interactive areas in playback videos, it would be very difficult to automate AR testing with them. 

VR testing tools (e.g., VRTest~\cite{vrtestdemo}, VRGuide~\cite{wang2023vrguide}) cannot be applied to testing with playback videos because they rely on the flexible control of camera movement and angle to explore the virtual scene, and their goal is to create the most efficient exploration paths of the camera. However, for playback videos, once they are created, the camera's exploration path becomes fixed, and the goal of testing tools becomes maximizing the interaction frequency and variety on the fixed exploration path.

In this paper, we present \textsc{TARIPlay}, a \underline{T}est framework for \underline{AR} applications based on \underline{I}nteractive area tracking in \underline{Play}back videos. In particular, \textsc{TARIPlay} addresses the special challenge of playback-video-based AR testing (compared with virtual-scene-based AR testing) by extracting test interfaces from videos. Given a Android ARCore playback video, our framework first uses the API of underlying ARCore session to extract all interactive areas within each frame of the video. Then, it uses a novel algorithm to combine interactive areas across frames and identify interactive areas that appear over a certain time period. After that, the framework filters the interactive areas with a number of criteria based on visibility, stability, and life span, to finally identify proper interactive areas as test opportunities. Information about these test opportunities are then fed to an automatic test engine to trigger UI events accordingly. Note that although our framework is mainly designed for automatic testing and we evaluate it accordingly, the information can also benefit test code developers so that they can tell when and where to trigger a UI event in their scripts.


We evaluated \textsc{TARIPlay} on four open source AR apps and nine playback videos which form the benchmark video set of ARTBank~\cite{artbank}. We compared \textsc{TARIPlay} with Monkey, which is the only existing tool applicable to AR app testing. Our evaluation shows that \textsc{TARIPlay} achieves average branch coverage of 55.80\%, compared with 41.98\% of Monkey. Our experiment also shows that, the number and quality of test opportunities extracted from a playback video are correlated with the test coverage on it, indicating that our extracted test opportunities can be used as a quality metrics for playback videos and filter out those of low quality and not suitable for testing.

In particular, our paper makes the following major contributions. 

\begin{itemize}
    \item An analysis and summarization of major challenges in playback-video-based AR app testing. 
    \item Novel techniques to identify test opportunities from playback videos by tracking the size, location stability, and life span of interactive areas. 
    \item An evaluation of our framework \textsc{TARIPlay} in the setting of automatic AR App testing, validating the strength of extracted test opportunities on enhancing code coverage (especially on AR interaction code) and evaluating playback videos for testing suitability.  
\end{itemize}



\vspace{-0.2cm}
\section{Background}
\vspace{-0.1cm}
In this section, we will introduce some background knowledge about the loading of AR playback videos into Android ARCore sessions, and the polygon transformation algorithms which we use for detecting interactive areas from video frames. We expect the information will help with better understanding of our approach. 
\vspace{-0.2cm}
\subsection{Playback Videos in ARCore Sessions}
\vspace{-0.1cm}
On an Android phone with ARCore support, playback videos can be recorded within an ARCore session using API methods \texttt{startRecording()} and \texttt{stopRecording()} of class \texttt{ARCoreSession}. The API methods will create an enriched MP4 video stream file which stores various data types, including the camera's video stream, phone sensor data, and analyzed environment information. The recorded playback videos can be loaded back into an ARCore session through API method \texttt{setPlaybackDatasetUri()}. After the loading, the underlying ARCore system will treat the videos as runtime camera inputs and any Android AR apps can be executed upon them. 

When a playback video is loaded into the ARCore session, we can use various API methods provided by ARCore to extract all detected trackables (interactive areas in the real world space, such as a plane or a human face) from the current frame of the video. A detected trackable is typically presented as a polygon, with the coordinates (in the real-world space with the camera as the origin) of its geometric center and boundary points available. \textsc{TARIPlay} is built upon these extracted data with follow-up data processing techniques to identify test opportunities.

\vspace{-0.2cm}
\subsection{Polygon Geometric Algorithms}
\vspace{-0.1cm}
Polygon geometric algorithms are a set of computer geometry algorithms used for solving questions about polygons. In our approach, we need to use a 3D vertex projection algorithm to map a polygon(i.e., a detected interactive area from the 3D space of the ARCore session) in real-world space to a 2D polygon on the phone screen plane. Since the mapped 2D polygon may go outside the boundary of the screen, we further need to use a polygon clipping algorithm to acquire the part of the polygon visible on the phone screen. 

The 3D vertex projection algorithm~\cite{hearn1997computer} maps 3D points from a virtual scene onto a 2D display. The algorithm rely on a camera projection matrix, which combines intrinsic parameters (like focal length and optical center, saved in the playback video) and extrinsic parameters (such as position and orientation) to transform 3D world coordinates into 2D screen coordinates. In our approach, we choose to use the 3D vertex project algorithm implemented in the ARCore framework to map polygon vertices to the phone screen. 

\begin{figure}
    \centering
    \includegraphics[width=0.7\linewidth]{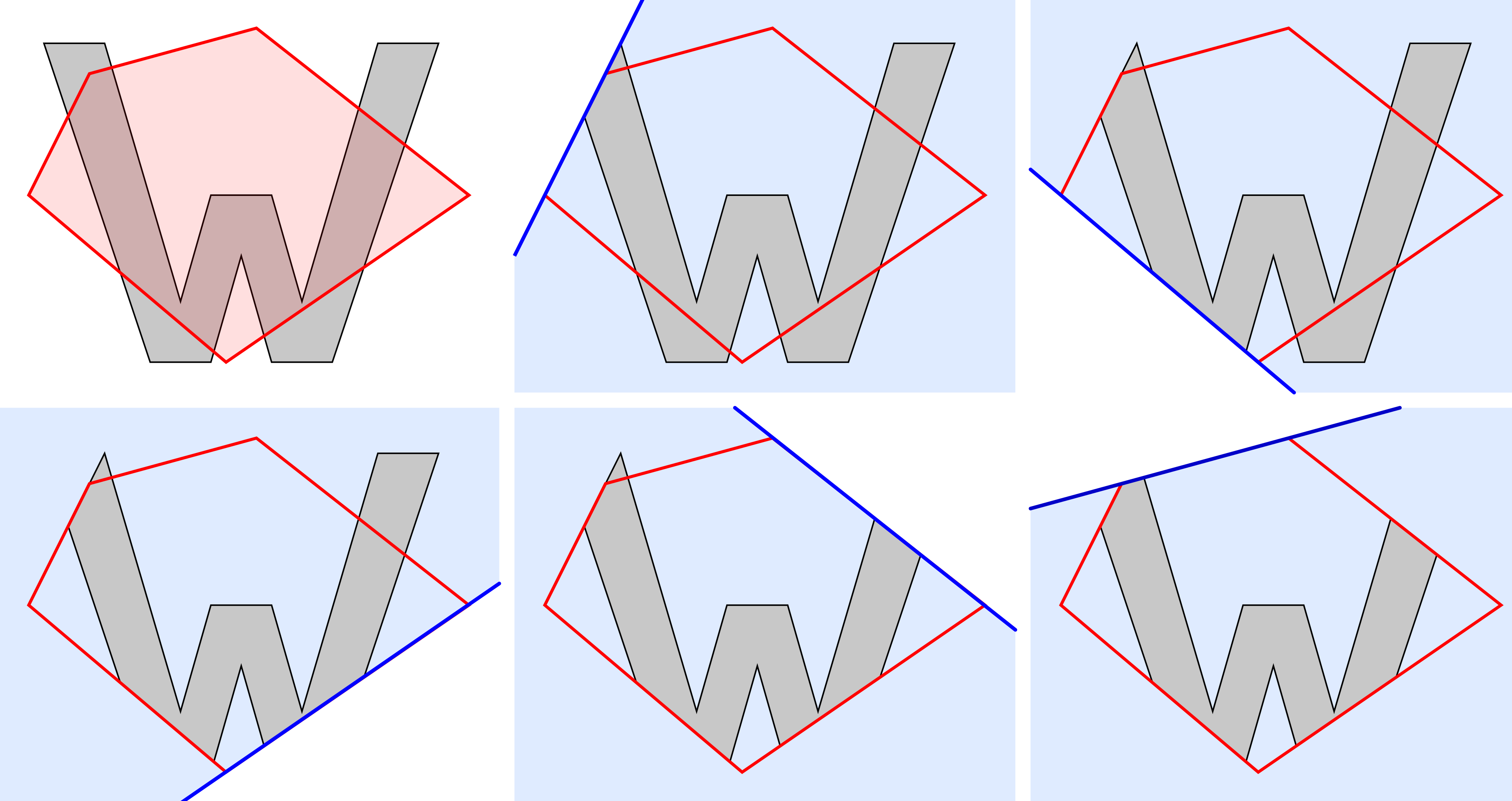}
    \vspace{-0.2cm}
    \caption{Illustration of Steps in Sutherland-Hodgman Algorithm (Top Left to Bottom Right)}
    \label{fig:suther}
    \vspace{-0.7cm}
\end{figure}

The Sutherland–Hodgman algorithm~\cite{sutherland1974reentrant} is a classic computer graphics algorithm used for polygon clipping, particularly to clip a polygon against a convex clipping window, such as a rectangle. For each edge of the clip boundary, the algorithm examines each pair of adjacent vertices in the polygon and determines whether the edge between them lies entirely inside, entirely outside, or partially inside the clip edge, and then generates new vertices accordingly. Figure~\ref{fig:suther} illustrates the algorithm on clipping a polygon ``W'' with a 5-sided polygon. 

\vspace{-0.2cm}
\section{Approach}
\vspace{-0.1cm}

\begin{figure*}
    \centering
    \includegraphics[width=0.9\linewidth]{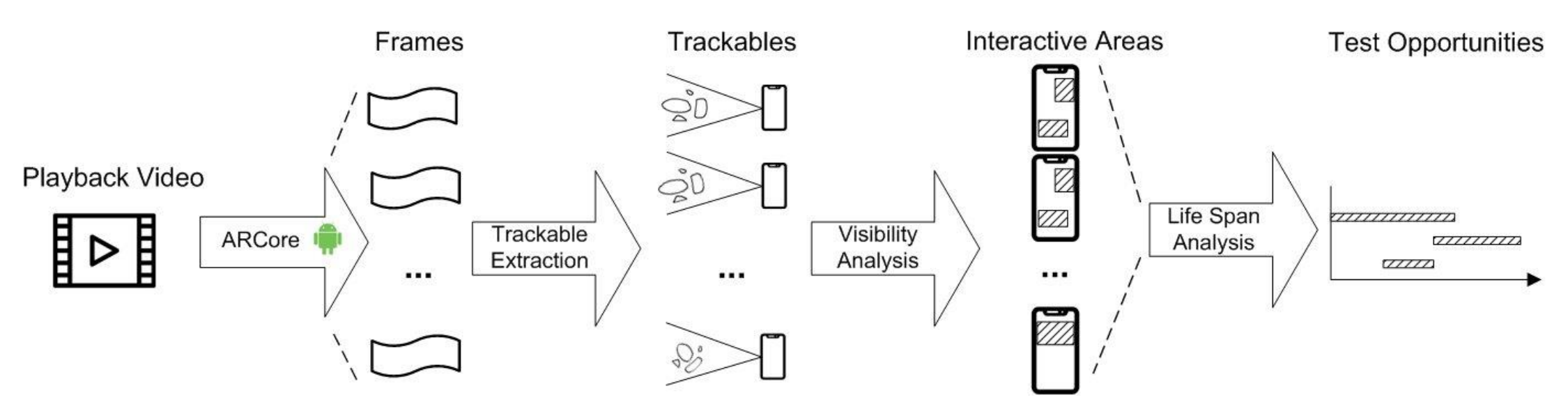}
    \vspace{-0.5cm}
    \caption{Overview of \textsc{TARIPlay}}
    \vspace{-0.5cm}
    \label{fig:overview}
\end{figure*}

In this section, we introduce our \textsc{TARIPlay} framework in details. \textsc{TARIPlay} takes a playback video as its input and identify a set of test opportunities (proper interactive areas with large enough size, stable position, and long enough life span). The overview of \textsc{TARIPlay} is presented in Figure~\ref{fig:overview}. From the figure, we can see that \textsc{TARIPlay} has four major steps: (1) using ARCore to split the playback video to a number of frames, (2) extracting trackables (interactive areas in the real-world space) from each frame, (3) performing visibility analysis to convert trackables to interactive areas on the phone screen, and (4) performing life span analysis to identify stable interactive areas on the phone screen and output them as test opportunities.

\vspace{-0.2cm}
\subsection{Video Loading and Frame Sampling}
\vspace{-0.1cm}
In the first step, we need to create an ARCore session to load the playback video. To achieve this, we create an empty Android AR app (as a part of \textsc{TARIPlay} whose only function is creating an ARCore session, loading a playback video and then extracting information about trackables from frames. \textsc{TARIPlay} also relies on ARCore's playback video loader to normalize frames to a standard aspect ratio (\(1920\times1080\)), ensuring consistency between different devices. The remaining analyses are performed offline after the playback video finishes. 


Since consecutive frames are very similar to each other, it is not necessary to analyze every frame. Therefore, \textsc{TARIPlay} samples frames at 10 frames per second (fps). We choose 10 fps as our sampling rate because $\sim$100 ms is a commonly reported lower bound for human visual recognition, so interactive areas with shorter lifetime are unlikely to be perceived \cite{vanrullen2001} \cite{thorpe1996}.  Our sampling reduces computational processing overhead and has an additional benefit to ignore some noises caused by erroneously rendered or missed single frame, yet we did not observe any loss of valid interactions in our experiments.



\vspace{-0.2cm}
\subsection{Trackable Extraction}
\vspace{-0.1cm}

From each sampled frame, \textsc{TARIPlay} extracts trackables available at the frame using ARCore API method  \texttt{session.getAllTrackables()}, and then we filter out trackables that are not tracked on the frame due to occlusions or tracking failures using below condition: \texttt{plane.getTrackingState() == TrackingState.TRACKING}. For each qualified trackable, \textsc{TARIPlay} collects the trackable ID, extents in all axes, and coordinates of the vertices and the center. Note that the trackable ID is used to link trackables across different frames. 

Based on the collected information, \textsc{TARIPlay} further filters out trackables not facing the camera so it is impossible to interact with them. Specifically, we filter out a trackable if the following condition is true: 

\vspace{-0.1cm}
\begin{equation}
\mathbf{n} \cdot (\mathbf{c} - \mathbf{p}) < 0
\end{equation}
\vspace{-0.1cm}

\noindent, where \(\mathbf{n}\) is the trackable’s normal vector, \(\mathbf{c}\) is the camera’s position, and \(\mathbf{p}\) is the trackable’s center.

\textsc{TARIPlay} currently considers only planes and filters out other types of trackables, because those trackables are not available in playback videos. Human faces are the second most commonly used trackables second to planes~\cite{verma20233}, but due to privacy concerns, few people are sharing playback videos with human face trackables, and we do not find such videos from the ARTBank~\cite{artbank} dataset. That being said, since human faces and other trackables such as pictures can also be modeled as polygons, our approach automatically applies to them as long as we do not filter them out. 

%




\subsection{Visibility Analysis}

After \textsc{TARIPlay} extracts all the trackables from a frame, it will perform visibility analysis to (1) calculate the projection of each trackable on the phone screen plane, (2) clip each projection area based on the phone screen scope and other overlapping projection areas, and (3) regularize the clipped projections from polygons to their inscribed rectangles. Using the inscribed rectangles as an approximation is an important design decision we make to simplify the technique. Although the inscribed rectangle may underestimate the interactive area of a trackable projection~\cite{PlaneRCNN}, this is a conservative approximation because UI events performed on the inscribed rectangle will be definitely received by the trackable. Furthermore, since testing tools typically send UI events to the center of an interactive area, the approximation will have only minimal influence on the testing process. On the other hand, using inscribed rectangle has large benefit as it allows easy and efficient calculation of area intersections, which is essential for the following life span analysis. 

The three steps in our visibility analysis are detailed below. 


\begin{itemize}
    \item \textbf{Trackable Projection}: To perform the projection, we first extract all polygon vertices from a trackable's representation, where each vertex \(\mathbf{v}_{\text{local}} = (x, y = 0, z, w = 1)^\top\) is defined in the plane's local coordinate system. Then, we project vertices to phone screen plane using the Model-View-Projection matrix chain:
        \[
        \mathbf{p}_{\text{screen}} = \mathbf{P} \cdot \mathbf{V} \cdot \mathbf{T} \cdot \mathbf{v}_{\text{local}}
        \]
    , where \(\mathbf{P}\) is the projection matrix, \(\mathbf{V}\) is the view matrix, and \(\mathbf{T}\) is the plane's pose transformation matrix. Note that we can acquire all the matrices from ARCore API methods.
    Finally, we convert \(p_{\text{screen}}\) to screen coordinates with Android Y-axis correction:
        \[
        \begin{aligned}
        x_{\text{screen}} &= \frac{(x_{\text{ndc}} + 1)}{2} \cdot W \\
        y_{\text{screen}} &= \left(1 - \frac{y_{ndc} + 1}{2}\right) \cdot H
        \end{aligned}
        \]
        where \(W\) and \(H\) are screen dimensions, and \((x_{\text{ndc}}, y_{\text{ndc}}) = (x_{\text{screen}}/w_{\text{screen}}, y_{\text{screen}}/w_{\text{screen}})\). After all vertices of a trackable polygon are projected to the phone screen plane, we can connect them to form the projection area.

    \item \textbf{Polygon Clipping of Projection Areas:} after projected areas are calculated, \textsc{TARIPlay} further performs polygon clipping (using Sutherland-Hodgman algorithm) to acquire the visible parts (i.e., areas that are within the scope of the phone screen and not occluded by other areas) of projected areas, which we refer to as \textit{visible polygons} later in the text. For clipping into the screen rectangle, we use the vertices vector \(\{(0,H), (W,H), (W,0), (0,0)\}\) in clockwise order. For each clipping edge, \textsc{TARIPlay} computes line intersections using parametric equations:
        \[
        t = \frac{(x_1 - x_3)(y_3 - y_4) - (y_1 - y_3)(x_3 - x_4)}{(x_1 - x_2)(y_3 - y_4) - (y_1 - y_2)(x_3 - x_4)}
        \]
        , where \((x_1, y_1)\) and \((x_2, y_2)\) define the polygon edge, and \((x_3, y_3)\) and \((x_4, y_4)\) define the clipping edge. To handle the occlusion, we use the same clipping approach. In particular, \textsc{TARIPlay} keeps track of distance of vertices to the camera, and ordered them from the closest to the farthest. For a given projection area, \textsc{TARIPlay} first identifies polygon edges (from all projection areas) that have at least one end point between the projection area's original plane and the phone, and then uses these edges to perform clipping on the projection area. Finally, the clipped area are removed and the resulted potentially concave area (if still existing) is partitioned to convex partitions. For efficiency purpose, we process the projection areas from the closest (to the phone) to the farthest. 
        
    \item \textbf{Calculation of the Inscribed Rectangle}: After all the visible polygons are acquired, \textsc{TARIPlay} further calculates their inscribed rectangles to facilitate later analyses. To achieve this goal, for a given visible polygon, it first extracts axis-aligned bounds from the polygon vertices as below, where (\(x_{\text{visible}}\), \(y_{\text{visible}}\)) is the coordinates of a vertex of the visible polygon. 
        \[
        \begin{aligned}
        x_{\text{raw}} &= \{\min(\mathbf{x}_{\text{visible}}), \max(\mathbf{x}_{\text{visible}})\} \\
        y_{\text{raw}} &= \{\min(\mathbf{y}_{\text{visible}}), \max(\mathbf{y}_{\text{visible}})\}
        \end{aligned}
        \]
        
    After that, for each vertex of the bounding box, we iteratively apply conservative shrinkage as below to shrink the bounding box to an inscribed rectangle. Note that a bound (e.g., (\(x_{\text{min}}\)) is moved until both vertices on its side (e.g., the two left vertices for \(x_{\text{min}}\)) go inside the polygon.  
        \[
        \begin{aligned}
        x_{\text{min}} &= \max(0, x_{\text{raw,min}} + 0.025 \cdot \Delta x) \\
        x_{\text{max}} &= \min(W, x_{\text{raw,max}} - 0.025 \cdot \Delta x) \\
        y_{\text{min}} &= \max(0, y_{\text{raw,min}} + 0.025 \cdot \Delta y) \\
        y_{\text{max}} &= \min(H, y_{\text{raw,max}} - 0.025 \cdot \Delta y)
        \end{aligned}
        \]
        where \(\Delta x = x_{\text{raw,max}} - x_{\text{raw,min}}\) and \(\Delta y = y_{\text{raw,max}} - y_{\text{raw,min}}\). After this process, we acquire an inscribed rectangle for each visible polygon, and we refer to these inscribed rectangles as \textit{visible boxes} later in the paper. The choice of 0.025 for bound reduction step is a tradeoff between performance and accuracy. We rely on Google ARCore to identify trackables and our on-the-fly analysis needs to keep up with the speed of video processing. Smaller steps improve accuracy but lead to more computation and sometimes video lags / crashes, while larger steps cause more inaccuracy of the calculated bounding box.
        

\end{itemize}

After \textsc{TARIPlay} completes the visibility analysis and acquires all the visible boxes for a frame, it will filter out all the visible boxes that are not large enough. We define \textit{Visibility Ratio} as below, and filter out all the visible boxes with \textit{Visibility Ratio} lower than 10\%. This default threshold is set according to the general guideline~\cite{nilsson2009design} for mobile UI design about the size of interactive areas which reliably support multi-finger interactions such as pinch and swipe. The movement and deformation of detected trackables in AR apps also call for larger interactive areas for stable interaction.

\[
\text{Visibility Ratio} = \frac{A_\text{box}}{A_\text{screen}} 
\]

, where \(A_\text{screen}\) is the screen area in pixels, and \(A_\text{box}\) is the area of the visible box in pixels. 



\vspace{+0.2cm}

\subsection{Life Span Analysis}
\vspace{-0.1cm}
While the size of visible boxes are important for effective testing, their life span is also important. If a visible box appears on the screen, and then disappears (due to lost tracking of the corresponding trackable) or moves outside the phone screen (due to movement of the phone during the video duration) in a short amount of time, it cannot be well leveraged in testing, because a UI event may not be successfully received due to computation delay, for multiple UI event sequences (e.g., tapping to create an object and then pinching to resize), visible boxes with short life span will cause the following event to be infeasible (i.e., the created object is no longer in the screen). Therefore, \textsc{TARIPlay} further calculate the life span of visible boxes and reports only those with long enough life span as test opportunities. 

\begin{figure}
    \centering
    \includegraphics[width=0.6\linewidth]{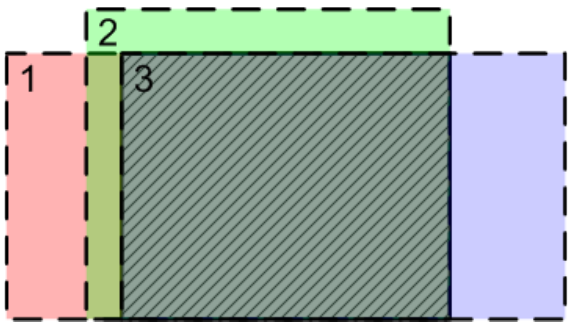}
        \vspace{-0.2cm}
    \caption{Visible Box Intersection in Life Span Analysis}
    \label{fig:life}
\end{figure}

The procedure of life span analysis is presented in Algorithm~\ref{alg:life}, and an illustration is shown in Figure~\ref{fig:life}. From the algorithm, we can see that, the life span analysis of a visible box starts at the first frame it appears in the playback video. Then, during following consecutive frames, the analysis continuously update the stable visible area and the last-visible frame. If visibility is no longer true (i.e., the stable visible area drops below 10\% of the screen) or the playback video ends, the algorithm will end and report all the frames between the starting and ending frames. From Figure~\ref{fig:life}, we can see that the stable visible area of a visible box shrinks to the box in the middle (marked with diagonal lines) as we intersecting the visible boxes at frames 1, 2, and 3, and that stable visible area is the safest screen area to send UI events during testing. 

\setlength{\textfloatsep}{8pt}
\begin{algorithm}[tb]
    \caption{Algorithm for Life Span Analysis}
    \label{alg:life}
    \hspace{-4cm}\textbf{Input}: VisibleBox, AllFrames\\
    \hspace{-4cm}\textbf{Output}: LifeSpan, StableBox\\
    \begin{algorithmic}[1] 
        \STATE Let $StableBox = FullScreen$
        \STATE Let $LifeSpan=\emptyset$
        \STATE Let $Started=False$
        \FOR{$Frame$ $\in$ $AllFrames$}
        \IF {$visible(VisibleBox, Frame)$}
            \STATE Let $Started=True$
            \STATE Let $StableBox=StableBox \cap VisibleBox$  
            \IF {$!visible(StableBox, Frame)$}
                \RETURN $StableBox, LifeSpan$
            \ENDIF
            \STATE Add $Frame$ to $LifeSpan$
        \ELSIF{$Started$}
        \STATE break
        \ENDIF
        \ENDFOR
        \RETURN $StableBox, LifeSpan$            
    \end{algorithmic}
\end{algorithm}

After we acquired life span of all visible boxes, we use the threshold of two seconds to filter out the visible boxes with life spans shorter than two seconds and report the remaining as test opportunities. We choose two seconds as the default threshold because it takes about 0.6 seconds for a human being to reflect on a visual stimulus and give hand response~\cite{gale1997human}~\cite{pfister2014comparison}, so two is the fewest number of whole second that allows two consecutive UI events to be triggered naturally. In addition, the sudden appearance of AR interactive areas and multi-finger gestures (gestures involving multiple finger movements within a single stable window) both require longer reflection time. Although we use automatic testing and coverage to evaluate TARIPlay, we believe that it is also important that the generated test cases simulate how typical users would interact with the app. Our experiments further studied the effectiveness of interactive area detection with different visibility area and life span thresholds. 


\subsection{Visualization and Interaction Scheduling}
\begin{figure}
    \centering
    \includegraphics[width=0.7\linewidth]{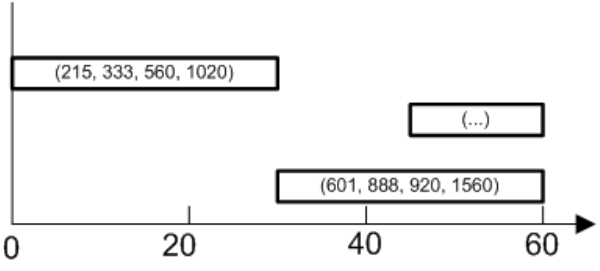}
        \vspace{-0.3cm}
    \caption{A Exemplar Gantt Chart Output of \textsc{TARIPlay}}
    \vspace{-0.3cm}
    \label{fig:gantt}
\end{figure}

To better present the test opportunities reported by \textsc{TARIPlay}, we can further visualize the visible boxes and their life spans in Gantt charts. For the example in Figure~\ref{fig:gantt}, we can see that visible boxes are presented as blocks along the time line of the video. If used for manual test code development, developers can utilize the chart to easily decide when to tap on a box and place a 3D object so that they have enough time later to interact with it. They may also easily identify the video period (e.g., second 45 to 60 in the figure) when multiple visible boxes exist so they can test cross-plane actions such as moving an object from one plane to another. It should be noted that, since Google ARCore's computer vision algorithms bring in randomness, the test opportunities created by \textsc{TARIPlay} may change when executed multiple times. Therefore, \textsc{TARIPlay} will analyze a playback video for three times and calculate the intersection of all test opportunities as its final output.

\subsection{Automated AR App Testing with \textsc{TARIPlay}}
While \textsc{TARIPlay} can be used to support human test developers to write tests more easily, it also provides a basis for automatic AR apps testing because it allows the test driver to always send UI events to interactive areas on the screen and thus makes the testing more efficient. It may also potentially enable the adaptation of traditional GUI testing strategies~\cite{said2020gui} to AR because the test opportunities provided by \textsc{TARIPlay} are comparable to the UI layouts provided by UIAutomator~\cite{uiautomator} (though the time constraints of test opportunities add more complexity and needs additional techniques to handle). 

In this paper, we will leave the above research opportunities for future and focus on the evaluation of the generated test opportunities themselves for automatic testing. Since Monkey~\cite{monkey} is the only automatic testing tool applicable to AR apps, for fair comparison, we implemented an automatic testing tool also with random strategy but on top of \textsc{TARIPlay} output and use it in our evaluation. In particular, our tool generates random UI events with the same distribution as Monkey. The difference is that, at a specific time point, we send the UI events only to coordinates within the visible boxes existing at the time point, according to the output of \textsc{TARIPlay}. For multi-coordinate gestures such as pinch, TARIPlay randomly chooses multiple coordinates within the identified interactive area.





\section{Evaluation}

We conducted an experiment study to evaluate the effectiveness of our random testing tool based on \textsc{TARIPlay} framework against Monkey, the existing automatic testing tool directly applicable to AR apps. 




\subsection{Research Questions}

The primary goal of our experiment is to find out whether our \textsc{TARIPlay} framework can enhance the effectiveness of automatic testing on AR apps. Therefore, we try to answer the following three research questions. 

\begin{itemize}
\item \textbf{RQ1:} What is the effectiveness of \textsc{TARIPlay} on detecting test opportunities in AR playback videos, with different thresholds of visibility area and life span?
    \item \textbf{RQ2:} How much does \textsc{TARIPlay} help to boost code coverage over Monkey on automatic AR app testing?
    \item \textbf{RQ3:} How does app complexity (e.g., gesture diversity, hybrid views) impact effectiveness of \textsc{TARIPlay} and Monkey?
    \item \textbf{RQ4:} How environmental factors and video qualities may affect achieved code coverage?
\end{itemize}

\subsection{Experimental Setup}

\textbf{Configurations of Compared Approaches.} Our experimental design employs a controlled comparison methodology using three distinct testing configurations: (1) the Standard Monkey configuration which executes Monkey testing with 50,000 completely random events and no throttling between actions, (2) the Enhanced Monkey which applies 50,000 events with 100ms throttling and restricts events to taps, swipes, and pinches (parameters: \texttt{--anyevent=0 --pct-majornav=0, --pct-appswitch=0, --pct-syskeys=0}) to approximate more realistic human interaction patterns and prevent excessive navigation events that would exit AR scenes prematurely, and (3) our proposed approach, which employs output of \textsc{TARIPlay} from playback videos, and strategically confines UI events to the detected visible boxes only during their life span. For UI events, we issue random sequence of tapping and gestures similar to Monkey but limit gestures to detected test opportunities. 




\textbf{Hardware Environment.} The experimental environment utilized a Xiaomi 11T Pro device running Android 14 with \texttt{testCoverageEnabled = true} and ProGuard/R8 disabled to ensure consistent hardware conditions and accurate instrumentation across all test runs. We acquire test coverage through instrumentation based on JaCoCo version 0.8.11 integrated with Gradle 7.3 and Android Debug Bridge version 34.0. Our analysis pipeline leveraged Python 3.11 scripts for statistical computation and visualization generation. 

\textbf{Subjects.} Our evaluation dataset comprises four open-source AR apps from Github and nine ARCore playback videos from the ARTBank benchmark folder~\cite{artbank}, forming 4 × 9 = 36 test settings. 

The subject open-source AR apps represent a diverse set of programming language and framework configurations (i.e., Java/Kotlin/C\# and ARCore/Sceneform/SceneView). Each app exhibits distinct interaction paradigms and complexity characteristics. This selection encompasses applications ranging from single-view AR experiences to hybrid interfaces combining traditional UI elements with AR components, supporting varying gesture vocabularies from simple tap interactions to complex multi-finger manipulations including rotation and scaling operations.

The ARTBank benchmark of nine playback videos represents diverse real-world testing conditions with balanced factor coverage: 3 indoor × 3 outdoor environments across 3 lighting conditions (low, medium, bright) with varying camera movement patterns from static positioning to high-motion scenarios. This stratified dataset allows us to assess our approach's robustness across heterogeneous deployment conditions commonly encountered in AR applications.


\subsection{Metrics and Analysis Framework}

Our evaluation employs multiple complementary metrics to comprehensively assess testing effectiveness. Code coverage analysis utilizes five JaCoCo metrics: instruction coverage measuring the percentage of executed bytecode instructions, branch coverage capturing decision path exploration, line coverage indicating source code reach, complexity coverage based on cyclomatic complexity paths, and method coverage tracking invoked functionality. Some AR apps also often have a GUI component to lead users to the AR sessions. Since our focus is on testing AR-related code and AR-specific behavior, we configure Jacoco to not considering Android framework boilerplate and restrict the coverage analysis to only application-specific packages containing AR-related code.

Beyond traditional coverage metrics, we introduce the Gesture Success Rate (GSR) as a novel metric specifically designed for AR testing evaluation. GSR quantifies the percentage of attempted gestures (UI input) that successfully trigger intended application responses, computed as the ratio of successful gesture completions to total gesture attempts on all trackables. Success is determined via deterministic application callbacks and logcat analysis that capture recognized gestures through temporal matching between triggered events and application logs. This metric proves particularly valuable for assessing multi-finger interaction reliability, where traditional random testing often fails due to spatial and temporal coordination requirements.


Because coverage distributions are non-Gaussian and we have repeated measurements of the same (app, video) scenarios across three methodologies, we use non-parametric repeated-measures procedures throughout. For each coverage metric (Line / Branch / Method), we run a Friedman test (blocks = (app, video) $\times$ run; treatments = \{Standard Monkey, Enhanced Monkey, TARIPlay\}). For pairwise contrasts we use Wilcoxon signed-rank tests with Holm family-wise adjustment across the three methodology pairs. We report Cliff's~$\delta$ (equivalently, paired rank-biserial) as the effect size with bias-corrected and accelerated (BCa) bootstrap 95\% CIs (10\,000 resamples). 



\subsection{Results and Analysis}

\subsubsection{RQ1: Detection of Interactive Areas.} To assess the quality of test opportunities detected by \textsc{TARIPlay} , we conducted a manual study on 9 playback videos. Two people (one of the authors and another student who is not an author) independently labeled test opportunities in the videos, so that the test opportunities detected by \textsc{TARIPlay} can be classified as \emph{opportunities} (labeled as a test opportunity by both people), \emph{borderlines} (labeled as a test opportunity by only one person), or \emph{non-opportunities} (labeled as a test opportunity by neither). The study results with different thresholds are presented in Table~\ref{tab:human-labeling}. We keep one threshold at its default value (2s or 10\%), and change the other threshold for larger (3s or 20\%) or smaller values (1s or 5\%). Columns 2-3 presents precision and recall when considering borderlines as opportunities (L for Loose), and Columns 4-5 presents precision and recall when considering borderlines as non-opportunities (S for Strict). 

The results show precision / recall of 79.7\% / 92.6\% (Loose) and 69.1\% / 96.3\% (Strict) using the default setting. Lowering duration/visibility thresholds yields marginal recall gains but noticeably reduces precision; increasing thresholds improves precision at the expense of recall (which is more important in testing), so the default setting (2s, 10\%) provides a good balance point. In practice, we observed two common false positive cases: (i) interactive areas that \emph{visibly shake} due to camera/scene motion, and (ii) \emph{very narrow} areas that are technically clickable but hard to interact with.


\begin{table}[t]
\centering
\caption{TARIPlay's effectiveness on recognizing Test Opportunities}
    \vspace{-0.4cm}
\label{tab:human-labeling}
\begin{tabular}{|l|r|r|r|r|}
\hline
\textbf{Threshold Setting} & \textbf{Prec.(L)} & \textbf{Rec.(L)} & \textbf{Prec.(S)} & \textbf{Rec.(S)}\\
\hline
2s+10\% & 79.7\% & 92.6\% & 69.1\% & 96.3\% \\
1s+10\% & 69.6\% & 95.4\% & 58.7\% & 96.3\% \\
3s+10\% & 86.6\% & 78.7\% & 76.3\% & 82.0\% \\
2s+5\%  & 66.6\% & 95.4\% & 55.8\% & 96.3\% \\
2s+20\% & 79.9\% & 81.0\% & 69.7\% & 83.6\% \\
\hline
\end{tabular}
\end{table}


\subsubsection{RQ2: Comparison of Coverage}

Table~\ref{tab:detailed_coverage_metrics} presents detailed code coverage comparison across all four applications, comparing Monkey baselines with our \textsc{TARIPlay} approach. We repeated each configuration three times and report the mean coverage in Table~\ref{tab:detailed_coverage_metrics}.  To quantify stochasticity, we also calculate across-execution variability over the three runs. For Monkey-S, the variance of line, branch, and method coverages are $\pm$4.21\%, $\pm$4.10\%, and $\pm$5.10\%, respectively. For Monkey-E, the variance of line, branch, and method coverages are $\pm$8.71\%, $\pm$8.58\%, and $\pm$9.58\%, respectively. For \textsc{TARIPlay}, the variance of line, branch, and method coverages are $\pm$0.54\%, $\pm$1.06\%, and $\pm$2.00\%, respectively. The lower variance for TARIPlay is consistent with its use of stable test opportunities.

\textsc{TARIPlay} achieves notable improvements across all metrics, with branch coverage showing substantial gains across all applications (+15.72 pp average over Standard Monkey, +13.82 pp over Enhanced Monkey). Method coverage demonstrates consistent improvements with an average of +12.25 pp over Standard Monkey and +10.62 pp over Enhanced Monkey, while line coverage shows meaningful improvement averaging +9.33 over vs Standard Monkey and +10.85 pp over Enhanced Monkey. 

\textbf{Statistical Analysis.} Our dataset comprises $4$ Apps $\times$ $9$ Videos $\times$ $3$ Executions = $108$ observations (per approach variant per metric). Omnibus Friedman tests show strong differences among methodologies for all three coverage metrics ($p < 10^{-13}$ for Line, Branch, and Method). For our primary contrast (TARIPlay vs.\ Enhanced Monkey), Wilcoxon signed-rank yields $p = 2.91 \times 10^{-11}$. Corresponding Cliff's~$\delta$ values indicate very large effects: $\delta = 0.890$ (Line), $0.849$ (Branch), $0.836$ (Method), all closing to 1.0. 



\begin{table*}[t]
\centering
\caption{Comprehensive Coverage Metrics Comparison}
    \vspace{-0.4cm}
\label{tab:detailed_coverage_metrics}
\small
\begin{tabular}{|l|l|c|c|c|c|c|}
\hline
\textbf{Application} & \textbf{Metric} & \textbf{Monkey S} & \textbf{Monkey E} & \textbf{TARIPlay} & \textbf{$\Delta$ (pp)} & \textbf{Improvement} \\
\hline
\multirow{3}{*}{AR Builder} & Branch & 42.44\% & 40.90\% & \textbf{58.18\%} & +15.74 & +37.1\% \\
 & Method & 71.25\% & 70.02\% & \textbf{80.42\%} & +9.17 & +12.9\% \\
 & Line & 66.53\% & 65.68\% & \textbf{77.78\%} & +11.25 & +16.9\% \\
\hline
\multirow{3}{*}{AR Ecommerce} & Branch & 45.00\% & 46.67\% & \textbf{49.63\%} & +2.96 & +6.3\% \\
 & Method & 71.43\% & 72.43\% & \textbf{81.07\%} & +8.64 & +11.9\% \\
 & Line & 67.57\% & 65.03\% & \textbf{69.37\%} & +1.80 & +2.7\% \\
\hline
\multirow{3}{*}{AR Simulator} & Branch & 36.96\% & 35.99\% & \textbf{58.45\%} & +21.49 & +58.2\% \\
 & Method & 58.62\% & 58.62\% & \textbf{70.50\%} & +11.88 & +20.3\% \\
 & Line & 53.27\% & 53.04\% & \textbf{68.34\%} & +15.07 & +28.3\% \\
\hline
\multirow{3}{*}{HelloAR} & Branch & 35.92\% & 44.35\% & \textbf{57.10\%} & +12.75 & +28.7\% \\
 & Method & 70.51\% & 77.25\% & \textbf{88.76\%} & +11.51 & +14.9\% \\
 & Line & 59.32\% & 66.83\% & \textbf{78.31\%} & +11.48 & +17.2\% \\
\hline
\multirow{3}{*}{Overall} & Branch & 40.08\% & 41.98\% & \textbf{55.80\%} & +13.82 & +33.0\% \\
 & Method & 67.95\% & 69.58\% & \textbf{80.20\%} & +10.62 & +15.1\% \\
 & Line & 64.17\% & 62.65\% & \textbf{73.50\%} & +9.33 & +14.5\% \\
\hline
\end{tabular}
    \vspace{-0.2cm}
\end{table*}

\begin{figure*}[h]
    \centering
    \includegraphics[width=0.95\linewidth]{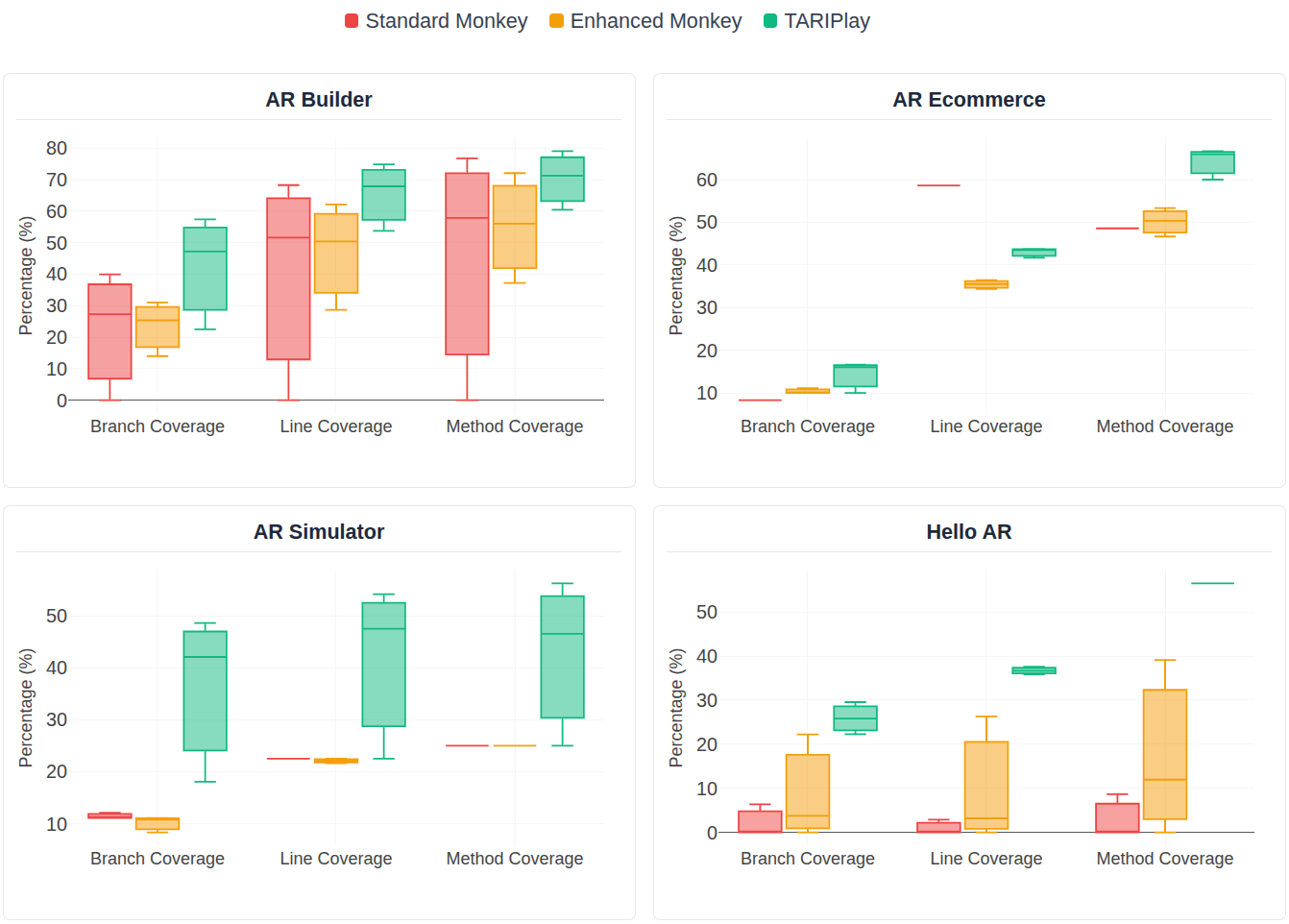}
    \Description{Omitted baseline from the coverage metrics}
    \vspace{-0.1cm}

    \caption{Coverage Comparison on Interaction Code}
\vspace{-0.1cm}

    \label{fig:ommit-baseline}
\end{figure*}





While we configure Jacoco to rule out non-AR-related code, there are also AR-related code which is executed all the time, including initialization of AR sessions, permission handling for storage and camera, resource loading, etc.  Compared with such code, it is more important to compare the coverage on code that can be triggered only with interactions in AR sessions. For each app, we identify always-executed code by running the app through each playback video without triggering any UI events.
When Jacoco provides code coverage at different code granularity levels, we remove the intersections between evaluated runs (Monkey variants and TARIPlay) and no-input runs.  The remaining code is considered \textit{interaction code} which is harder to cover and more important on comparing effectiveness of testing techniques.  Figure~\ref{fig:ommit-baseline} further compares code coverage on the interaction code.  

Figure~\ref{fig:ommit-baseline} is a group of four box plot (one plot for each app) where each box describes the distribution of interaction code coverage over different playback videos in the corresponding setting. From the figure, we have two observations. First, for each app, our approach outperforms both Monkey variants on almost all videos, showing consistent enhancement also over different videos. Second, our approach achieves a larger coverage improvement on interaction code, show over 20 percentage point gain in all apps except for AR Ecommerce. The figure also shows variance on absolute code coverage and distributions across different apps and coverage criterion, which we detail below.




\textbf{Hello AR:} This app supports placing and removing of multiple objects. \textsc{TARIPlay} performed well consistently on the app. While Monkey variants sometimes can achieve acceptable coverage (on videos with abundant test opportunities), the variance is very large as they cannot stably trigger AR interactions. 


\textbf{AR Ecommerce:} This application allows single object placement (\texttt{model.setParent(anchorNode)}) and maintains object selection for following user interactions with the placed object. The variance is low for all approaches because just catching one test opportunity will allow successful placement of objects. \textsc{TARIPlay} performed still better because it restricted the following interactions within the interactive area, and thus boosted their success rate.




\textbf{AR Builder:} This application supports tap, move, and scaling gestures but not rotation. It provides semi-AR functionality including rescaling, object removal, screenshots, color/type selection through popups, and input fields for object dimensions. Monkey performs better in ARBuilder than other apps (still with high variance due to its high randomness) because it supports more Semi-AR (traditional GUI inside AR) features.


\textbf{AR Simulator:} This app features placement of a single object and supports dragging through AR gestures for precise movements, plus rotation capabilities similar to AR Builder but limited to one placed object per session. Our approach performs much better than Monkey because it limits gestures into visible boxes so more complicated gesture sequences in the same area are more likely to happen. Our performance varies across videos because some videos do not have enough test opportunities for complicated gestures (e.g., dragging for precise movements) supported in the app. 


\subsubsection{RQ2: Coverage vs Complexity}

Results in Table~\ref{tab:detailed_coverage_metrics} and Figure~\ref{fig:ommit-baseline} show that our approach makes larger improvements on the two more complicated apps AR Builder and AR Simulator. Statistical analysis reveals relations between application gesture complexity. AR Builder, supporting four distinct gesture types including complex multi-finger operations, shows the largest overall coverage improvements. Hello AR, with its simplified interaction model focused primarily on tap gestures, demonstrates consistent but more modest gains. This pattern suggests that our approach provides greatest value for applications implementing sophisticated interaction paradigms that challenge existing testing tools.



To better understand the performance different techniques on more complicated gestures, we performed gesture success rate analysis and compare the gesture success rate of different techniques on the four apps. The results are presented in Table~\ref{tab:gesture_success_rates}. 

From the table, we can see substantial improvements in multi-finger interaction reliability when comparing \textsc{TARIPlay} against Monkey. 
While baseline Monkey achieves acceptable success rates for simple tap gestures at 94.3\%, performance degrades substantially for complex interactions. In particular, for Monkey, Drag gesture success rates reach only 39.7\%, rotation gestures succeed in merely 34.2\% of attempts, and scaling operations achieve only 28.9\% success rates. Analysis of Monkey's limitations reveals three key challenges: (1) \textbf{Spatial Ignorance}—Monkey touches are completely random across the screen without knowledge of placed object locations or anchor positions, failing to target specific anchors for manipulation and lacking understanding of AR spatial relationships; (2) \textbf{Chaotic Movement Patterns}—Monkey generates unpredictable ACTION\_MOVE sequences that may not maintain proper single-finger contact or satisfy hit-testing requirements; (3) \textbf{Validation Criteria Failures}—Even when near anchors, Monkey's erratic movement patterns may not meet the validation criteria needed for gesture recognition. The challenges are compounded when applications deselect placed objects after placement (Hello AR vs AR Simulator), as 3D raycasting constants (3D bounding spheres or 2D constants) may degrade gesture support if bounding spheres are insufficiently large for normal interaction requirements.

\textsc{TARIPlay} addresses these limitations by precisely aligning multi-finger events with confirmed stable plane regions and appropriate temporal windows. In particular, drag gesture success rates improve to 92.4\%, while rotation and scaling operations achieve 86.8\% and 89.3\% success rates respectively. Overall gesture success rate increases from 67\% with Monkey to 96\% with our approach.

\begin{table}[t]
\centering
\caption{Gesture Success Rates by Type and Application}
    \vspace{-0.3cm}
\label{tab:gesture_success_rates}
\small
\begin{tabular}{|l|c|c|c|c|}
\hline
\textbf{Type} & \textbf{AR Builder} & \textbf{AR Ecomm.} & \textbf{AR Simul.} & \textbf{HelloAR} \\
\hline
\multicolumn{5}{|c|}{\textbf{Monkey}} \\
\hline
Tap & 96\% & 98\% & 99\% & 97\% \\
Drag & 41\% & 45\% & 48\% & 42\% \\
Rotation & 32\% & 39\% & -- & 35\% \\
Scaling & 28\% & 34\% & -- & 29\% \\
\hline
\multicolumn{5}{|c|}{\textbf{TARIPlay}} \\
\hline
Tap & 100\% & 100\% & 100\% & 100\% \\
Drag & 95\% & 92\% & 96\% & 94\% \\
Rotation & 91\% & 88\% & -- & 86\% \\
Scaling & 93\% & 89\% & -- & 92\% \\
\hline
\end{tabular}
\end{table}

\subsubsection{RQ3: Video Factor Impact}

\begin{figure}[h]
    \centering
    \includegraphics[width=\linewidth]{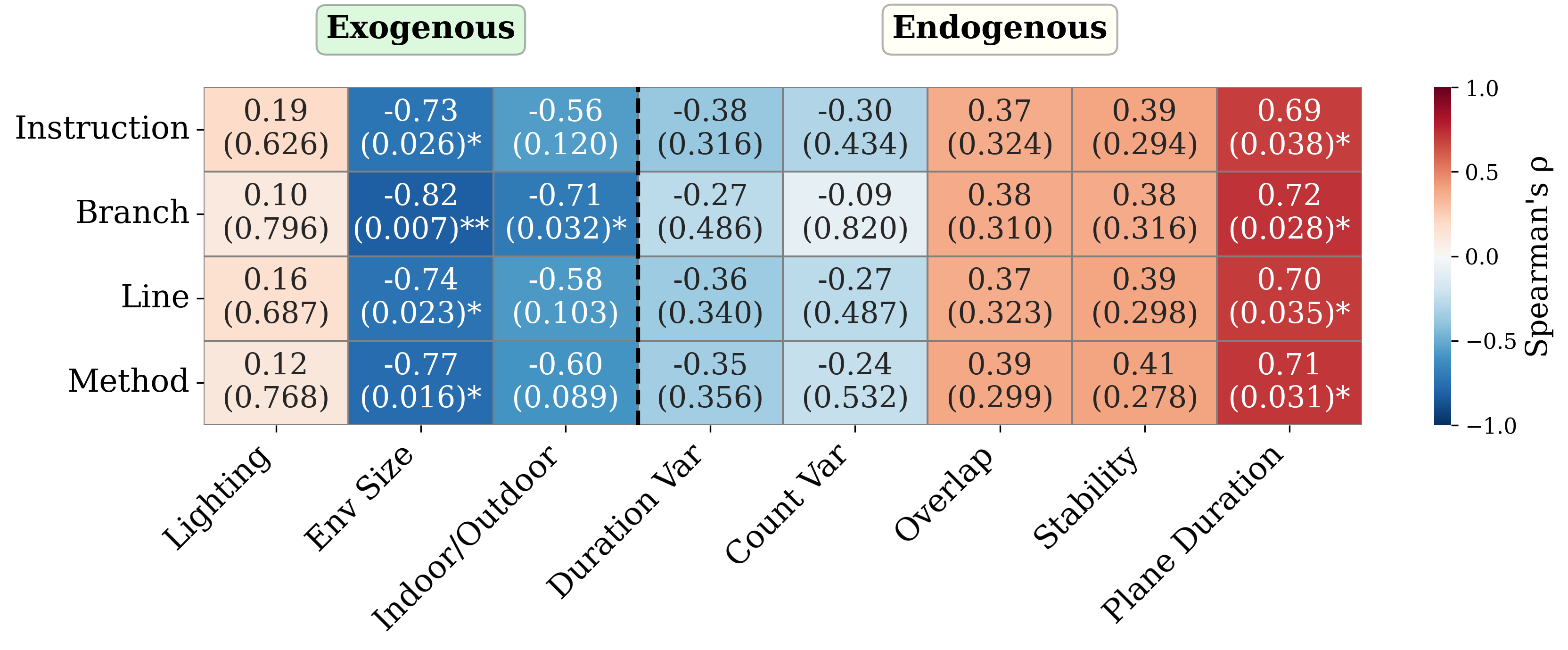}
    \Description{Omitted baseline from the coverage metrics}
        \vspace{-0.4cm}
    \caption{Correlation Between Coverage and Video Factors, *p<0.05, **p<0.01}
        \vspace{-0.3cm}
    \label{fig:heatmap}
\end{figure}

Figure~\ref{fig:heatmap} shows the correlation between achieved code coverage and various factors from the playback videos. Because environmental variables are ordinal, we use Spearman’s rank correlation ($\rho$). For each cell we compute the p-value as shown in the brackets. 
Exogenous variables (to the
left of the dashed line in figure ~\ref{fig:heatmap}) are environment factors related to general video features, whereas variables on the right are TARIPlay-derived metrics. Among the endogenous variables, average plane duration (average lifespan of visible boxes) has the largest observed positive correlation with branch coverage ($\rho$ $\approx$ 0.73, p $\approx$ 0.03), suggesting that videos providing longer stable interaction opportunities enable more comprehensive testing. Mutual stability and mean overlaping areas show similar positive trends. Since Google ARCore's computer vision algorithms bring in randomness, the test opportunities created by \textsc{TARIPlay} may change when executed multiple times. After three executions, we calculate the result similarity as mutual stability, and mean overlapping areas as the average overlapped test opportunities across executions. 
We can see that when the extracted test opportunities are more stable across executions, the more likely we can achieve a higher code coverage. However, none of these correlations are statistically significant at $p=0.05$, so we consider these relationships as descriptive patterns rather than definitive correlations.

For exogenous environmental factors, we convert lighting conditions (low = 1, hight = 3), environment sizes (small = 1, large = 3), and indoor / outdoor (indoor = 1, outdoor = 2) to numbers and performed the correlation analysis (the labels are provided by ARTBank). From the figure, we can see that lighting condition is slightly correlated with code coverage, but both environment sizes and indoor / outdoor settings have consistently negative $\rho$ values, showing that it is more difficult to achieve high coverage in playback videos of larger spaces and outdoor settings. 



\subsubsection{UI Events Statistics}

\textsc{TARIPlay} demonstrates improvements on the number of UI events triggered compared to Monkey approaches while maintaining superior coverage effectiveness. We set the total number of events (per app per video per execution) to 50,000 to maximize Monkey's test coverage potential, and 50,000 was never reached before the video stops. Baseline Monkey with no throttle triggers 18,171 events per execution on average . Enhanced baseline with 100ms delays between events triggers 2,182 events per execution on average, while TARIPlay triggers only 198 event per execution on average. Please note that although all executions take exactly the same time (video length), triggering fewer UI events indicates that \textsc{TARIPlay} tests are more meaningful and may be adapted to human tests more easily. 


\subsection{Threats to Validity}
The major threat to construction validity is the coverage-quality assumption that higher code coverage always leads to improved fault detection capability, and it may not be the case~\cite{kochhar2015code}. In AR apps, a same UI events on different trackables may trigger exact the same code, but have different visual effects, so sending more UI events to cover more trackables is also important. To mitigate this threat, we supplemented coverage analysis with gesture success rates to provide multiple effectiveness perspectives. The major threat to internal validity is the instrumentation bias where JaCoCo coverage measurement may miss dynamically-loaded methods or reflection-based code execution. To mitigate this threat, we manually verified the class loader runtime logs and confirmed that all the loaded classes are from the original source code package the corresponding executions are recorded in the coverage reports. Another threat to internal validity is the randomness in evaluation results caused by non-determinism in \textsc{TARIPlay}. To mitigate this threat, we run three executions for both the test opportunity detection and the app testing. For further mitigation, we plan to run more executions and investigate how many executions are sufficient for the evaluation of AR testing referring to the guidelines for randomized algorithm in testing \cite{10.1145/1985793.1985795}.
The major threat to the external validity is that our evaluation is limited to the Android ARCore framework, the Android ARCore Apps / playback videos , or just to the subject apps / playback videos / mobile device used in the experiment. Therefore, our findings may not be generalized to other apps / videos and Apple ARKit and its corresponding playback videos. To mitigate this threat, we used  four apps that cover multiple AR domains and interaction styles. We include single-view, pure-AR experiences alongside AR with extensive UI components. Interaction styles range from tap-only object placement to persistent multi-finger object engagement. The apps span retailing, education, and simulation domains. The nine videos are the benchmark provided in ARTBank and were chosen to cover different scene sizes, indoor/outdoor settings, and lighting conditions. Correlation and effect-size claims on a relatively small dataset (4 apps $\times$ 9 videos) risk low power and inflated Type-I error under multiple comparisons. For coverage comparisons we use non-parametric repeated-measures tests with Holm correction and report distribution-free effect sizes with uncertainty. In the future, we plan to mitigate this threat by performing experiment on more apps / playback videos and ARKit apps / playback videos. 






\subsection{Replication Package}

To support reproducibility and facilitate adoption by the AR testing community, we provide a comprehensive replication package\footnote{ https://anonymous.4open.science/r/TARIPlay-53F4} including all experimental artifacts and analysis infrastructure. Our package contains \textsc{TARIPlay}'s source code and subject apps / playback videos. It also contains complete Python analysis scripts for statistical computation and visualization generation, raw JaCoCo coverage reports and gesture success logs from all experimental runs, and detailed reproduction instructions with parameter configuration files. The replication package includes complete video corpus with extracted metadata in structured JSON format, enabling validation of our mutual interval computation approach and exploration of alternative scheduling strategies.

\section{Discussion}
\subsection{Playback Videos vs. Simulated Scenes}
Playback videos enable AR user interaction (e.g., object placement, object manipulation) with a pre-recorded video of a physical environment. One limitation of testing with playback videos is that they do not allow dynamic configuration of the testing scene (e.g., adjusting position of physical objects in the scene or changing lighting conditions) because the videos can hardly be revised after their generation. We believe this limitation can be largely alleviated when large sets of playback videos are accumulated and shared on the Internet, allowing AR testers to choose a subset of videos that cover the usage scenarios of their apps. Note that the required cost and expertise of creating playback videos are very low and even a non-technical person can create a lot of them with a recorder app. 

An alternate approach is to use a simulated physical environment (i.e., a VR scene) as the testing scene, which allows more flexible dynamic scene configuration. It will also allow conditional test cases (e.g., moving the camera based on intermediate test output). However, more studies are required to understand how the imprecision of the simulated environment may affect AR testing. For example, the mathematically generated surface may hide bugs related to imprecise plane detection, which is common for rough surface in the real world. Unnatural lighting conditions and reflections may also cause rendering difference of virtual objects, which may either miss rendering errors or cause false positives in testing. As for now, simulated environments are also supported in mainstream AR frameworks such as in Unity Mars~\cite{unitymars} and Google ARCore Virtual Scene~\cite{arcore}. As they complement each other, we envision that both playback videos and VR test scenes will play important roles in automated AR testing. 

\subsection{Adapting Existing GUI Testing Strategies}

After \textsc{TARIPlay} identifies test opportunities from playback videos, it can provide a UI layout at each time point similar to UIAutomator~\cite{uiautomator}, so it becomes possible to apply existing Android GUI testing strategies to test AR apps. For example, we can use model-based testing~\cite{baek2016automated} to trigger events within a visible box or across multiple visible boxes based on state transitions. We may also use pattern-based testing~\cite{costa2014pattern} to trigger a UI event sequence pattern on a visible box, hybrid analysis~\cite{zhang2021condysta} to take advantage of static information and combinatorial testing~\cite{niu2018interleaving} to exhaust different interaction combinations. However, even with support of \textsc{TARIPlay}, the still-existing unique challenges of AR testing are time constraints (events must be finished within time limits) and the lack of back button (one can never go back to a previous state because the playback video always goes forward). So more advanced scheduling algorithm is required to better order UI events and distribute them into different time slots. 




\section{Related Works}
Our research is mainly related to the testing of AR/VR applications, and the temporal tracking of planes. 

\subsection{Testing and Analysis of AR and VR Apps}
Two recent works PredART~\cite{Rafi2022} and VOPA~\cite{yang2024towards} facilitates AR app testing, but they focus on predicting test oracles to automatically detect AR bugs in screenshots or videos recorded during testing. In another related area of autonomous driving, Stocco et al. developed SelfOracle~\cite{9284027}, a self-assessment framework that monitors deep neural network confidence at runtime to predict unsafe or unsupported driving scenarios. These techniques can complement \textsc{TARIPlay} by checking its output videos after test interfaces are detected and interactions are performed. On VR testing, AutoWalker~\cite{autowalk} is similar to a Monkey for VR and randomly guides the player camera in the VR scene. Later, Wang et al. developed VRTest~\cite{vrtestdemo} and VRGuide~\cite{wang2023vrguide} to more efficiently explore VR spaces by detecting shorter paths in the space to interact with virtual objects. AutoWalker, VRTest, and VRGuide perform testing by controlling camera movement within interactive virtual scenes, which is not possible for fixed-camera playback videos. More recently, Li et al. developed a series of techniques~\cite{li2024less,li2025extended} to study and automatically assess cybersickness in virtual reality applications. On VR game testing, Gil et al.~\cite{gil2020automated} and Souza el al.~\cite{correa2018automated} proposed approaches to model VR applications and cover the model nodes and edges using automatic test cases. Zhao et al.~\cite{zhao2022lightweight} developed an testing approach to learn from human player behaviors.

Rzig et al.~\cite{rzig2023virtual} studied the characteristics of unit tests in VR applications and found they were of lower quality than their counterparts in other applications. It should be noted that testing of AR apps has very different challenges than the testing of VR apps. First, AR apps need to be tested in real-world scenes which are difficult to set up and reproduce, so playback videos and \textsc{TARIPlay} become necessary. In contrast, VR apps can be directly tested within the VR scene. Second, VR apps have all interactive objects predefined in the code, but AR apps read interactive objects from a changing environment so it always face issues of instable irregular interaction interface.  Vision-based testing~\cite{yu2023vision,xiao2019iconintent} is another promising approach that may be applicable to AR apps, but existing techniques focus on static screen images instead of videos. Note that \textsc{TARIPlay} is already built upon the output of ARCore, leveraging the powerful computer vision techniques from Google. 


Several empirical studies have investigated VR and video game software. Murphy-Hill et al.~\cite{murphy2014cowboys} examined video game developers to identify the unique challenges they face compared to traditional software development. Washburn et al.~\cite{gamestudy} analyzed failed game projects to uncover common pitfalls, while Lin et al. explored update patterns on the Steam platform to determine the priorities of game updates. Rodriguez and Wang~\cite{rodriguez2017empirical} studied the characters and trends of open source XR projects. Pascarella et al.~\cite{pascarella2018video} looked into open-source video game projects to highlight their characteristics and how they differ from non-game software projects. Harms~\cite{harms2019automated} introduced a set of guidelines for evaluating the usability of AR applications and classified various usability issues. Nusrat et al.~\cite{nusrat2021developers} examined performance repair logs of VR applications to pinpoint the main causes of performance degradation. Zhang et al.~\cite{zhang2019privacy} studied the privacy risks of AP apps in mobile computing context~\cite{zhang2020does,slavin2016toward,wang2018guileak}. Molina et al.~\cite{molina2021automatic} developed a code dependency analysis to detect dependencies across code and VR assets.




\subsection{Temporal Tracking Methods based on Stability}

Since our work tries to identify stable areas in a space over time, it is related to the series of work in AR to identify trackables based on their stability. Temporal tracking in AR involves keeping consistent identification of the same plane across multiple frames, and it is supported in ARCore. ARCore provides unique plane trackables (with persistent hash IDs within a session) that can be used to follow a plane over time~\cite{google2025arcoreplane}. Our approach also leverages these trackable IDs to avoid expensive re-detection. 

There are also works on assessing plane stability over time. Some recent research \cite{9417783} measure how consistent the plane’s position and shape remain across the selected frames. One method~\cite{PlaneRCNN} is to compute the overlapping area of the plane’s footprint between successive frames – for example, using an Intersection over Union (IoU) metric as is done in . A high overlap ratio over time suggests the plane region is stable~\cite{iou_metric}. Frame skipping~\cite{frame_skipping} is a technique to improve processing speed, with the trade-off of potentially missing rapid changes. Although these works also try to measure stability of planes in the 3D space, our approach works on a different setting (playback videos) and has a different purpose (focusing on AR app testing), so we developed visibility and life span analyses based on projection, and created different criteria (trackables are considered stable as long as their stable visible area is larger enough). 



\section{Future Works}
In the future, we plan to work on the following research directions. 
First, we will extend our experiment to include more apps and playback videos, studying the effectiveness of our approach in more experiment settings. Second, we will extend our framework to support additional AR platforms such as ARKit and Unity AR Foundation and study whether our approach can be also applicable to a different AR framework. Third, we plan to study how much the test opportunities detected by \textsc{TARIPlay} can support manual test script development and evaluate the usefulness of the produced Gantt charts for AR testers. In particular, we can ask human subjects to write AR test code for playback videos with and without \textsc{TARIPlay}, and compare their working efficiency. Fourth, \textsc{TARIPlay} currently does not consider the occlusion among placed virtual objects and trackables, which may lead to unusable test opportunities. We plan to develop techniques to filter occluded interactive areas. Finally, we will work on the adaptation of existing GUI testing strategies to AR apps based on \textsc{TARIPlay} output, developing new techniques with consideration of time constraint and optimized event scheduling. 


\section{Conclusion}

This work addressed the fundamental challenge of effective automated testing in Augmented Reality applications, where traditional random testing approaches fail to account for the spatial and temporal constraints inherent in AR interactions. We developed \textsc{TARIPlay}, a metadata-guided testing framework that leverages stable plane detection to schedule UI interactions within optimal spatio-temporal windows. Our empirical evaluation across four AR applications and nine representative testing scenarios demonstrates substantial improvements over existing approaches. \textsc{TARIPlay} achieves an average branch coverage improvement of 13.82 percentage points (from 41.98\% to 55.80\%) and gesture success rate improvements of 41.7 percentage points (from 51.8\% to 93.5\%). These gains stem from our method's spatial awareness and temporal coordination capabilities, which enable systematic exploration of AR-specific code paths that random testing frequently misses\blfootnote{This work is supported in part by NSF grants CCF-1846467, CCF-2007718, CNS-2221843, and CCF-2418093. We also would like to thank Sahan Kalutarage for helping to label ground-truth test opportunities in playback videos.}. 



\bibliographystyle{ACM-Reference-Format}
\bibliography{references}

\end{sloppypar}

\end{document}